\documentclass[aps,pra,twocolumn,superscriptaddress,amsmath,amssymb,showpacs]{revtex4}
\usepackage{graphicx}
\usepackage{amsmath}
\usepackage{amsfonts}
\usepackage{amssymb}
\usepackage{hyperref}
\begin{document}

\title{Equation of state and phase transition in spin-orbit-coupled Bose gases at finite temperature: A perturbation approach}

\author{Zeng-Qiang Yu}
%\date{\today}

\affiliation{Dipartimento di Fisica, Universit\`{a} di Trento and INO-CNR BEC Center, I-38123 Povo, Italy.}

\begin{abstract}
    We study two-component Bose gases with Raman induced spin-orbit coupling via a perturbation approach at finite temperature. For weak coupling, free energy is expanded in terms of Raman coupling strength up to the second order, where the coefficient (referred to as Raman susceptibility) is determined according to linear response theory. The equation of state for the stripe phase and the plane-wave phase are obtained in Popov approximation, and the first order transition between these two phases is investigated. As temperature increases, we find the phase boundary bends toward the stripe phase side in most temperature regions, which implies the ferromagnetic order is more robust than the crystalline order in presence of thermal fluctuations. Our results qualitatively agree with the recent experimental observation in rubidium atomic gases. A method to measure Raman susceptibility through the two-photon Bragg scattering experiment is also discussed.
\end{abstract}

\maketitle

\section{Introduction}

The remarkable realization of synthetic spin-orbit coupling in quantum gases is opening new perspectives in the study of many-body phenomena with ultracold atoms~\cite{Review}. To date, a specific type of spin-orbit coupling, which is generated by a pair of Raman laser beams, has been experimentally achieved in a Bose-Einstein condensate (BEC) of $^{87}$Rb~\cite{NIST} and degenerate Fermi gases of $^{40}$K~\cite{ZhangJing} and $^{6}$Li~\cite{MIT}.
For bosonic atoms, Raman coupling could give rise to novel superfluid phases due to the dramatic modification of the single-particle dispersion~\cite{NIST,Ho,Trento1}. One particularly interesting phase is the stripe (STR) BEC, which simultaneously breaks the U(1) symmetry and the translational invariance symmetry. In current experiments with Rb gases~\cite{NIST,ChenShuai}, due to the slight difference between inter- and intra-species interactions, the stripe phase is expected to exist only in the weak Raman coupling regime~\cite{Ho,Trento1}. At a critical Raman strength, a first order transition between the STR phase and plane-wave (PW) phase takes place~\cite{NIST,Ho,Trento1}. Since the PW condensate is almost fully-polarized in this regime, the STR-PW transition can be also regarded as a consequence of the competition between the crystalline order and the ferromagnetic order.

So far, the spatial density modulation in the stripe phase has not been directly observed experimentally. The major difficulties are the small contrast of the stripes and the limited resolution of {\it in-situ} imaging. Nevertheless, a miscible-immiscible transition has been identified at a value of the Raman strength \cite{NIST}, which is close to the theoretically predicted critical point of the STR-PW transition. Very recently, the temperature dependence of the phase boundary is experimentally determined through the statistical analysis of magnetization measurements \cite{ChenShuai}.

Previously, most theoretical works focused on the ground state properties and quantum dynamics of the STR phase and the PW phase~\cite{Ho,Trento1,Trento2,Trento3,Trento4,ZhengWei}; very few studies pay attention to the  thermal effects at finite temperature~\cite{ZhengWei}. In particular, to our knowledge, the influence of thermal fluctuations on the STR-PW transition has not been addressed in the literature.  In this work, we develop a perturbation approach to resolve this problem in the weak Raman coupling regime. By expanding the equation of state in terms of the Raman strength up to the second order, we find a temperature dependent STR-PW phase boundary, which is in qualitative agreement with the recent experimental observations~\cite{ChenShuai}.

\section{Perturbation formalism}

Consider a two-component Bose gas coupled by a pair of counter propagating Raman beams along the $x$ direction, the effective Hamiltonian is given by
\begin{gather}
   \hat H = \hat H_0 + \Omega \hat R + \hat H_{\rm int}, \label{Hamiltonian-1} \\
   \hat H_0 = \int d{\bf r}\, \hat \psi^\dagger \left(-\tfrac{\hbar^2\nabla^2}{2m} \check 1 - h \check \sigma_z \right) \hat \psi, \label{Hamiltonian-2} \\
   \hat R = \tfrac{1}{2} \int d{\bf r}\, \hat \psi^\dagger \left( \check \sigma_+ e^{-i2k_{\rm r} x} + {\rm H.c.} \right) \hat \psi, \label{Hamiltonian-3} \\
   \hat H_{\rm int} = \tfrac{1}{2} \sum_{\sigma,\sigma'=\uparrow,\downarrow} g_{\sigma\sigma'} \int d{\bf r}\, \hat \psi_\sigma^\dagger \hat \psi_{\sigma'}^\dagger \hat \psi_{\sigma'} \hat \psi_\sigma \label{Hamiltonian-4}
\end{gather}
where $\hat\psi_{\sigma}$ is the field operator for the pseudo-spin $\sigma$, $\hat \psi^\dagger = (\hat \psi_\uparrow^\dagger, \hat \psi_\downarrow^\dagger) $, $m$ is the mass of atoms, $k_{\rm r}$ is the recoil wave vector of the laser beams, $\Omega$ is the Raman coupling strength, $h$ is the effective Zeeman field fixed by the Raman detuning, $g_{\sigma\sigma'}$ are the contact interaction parameters with $g_{\uparrow\downarrow}=g_{\downarrow\uparrow}$, $\check\sigma_{x,y,z}$ are Pauli operators, $\check\sigma_{\pm}=\tfrac{1}{2}(\check\sigma_x\pm i\check\sigma_y)$, and $\check 1$ is the identity matrix.

When the Raman coupling is weak, one can treat $\Omega \hat R$ as a perturbation and expand free energy in terms of $\Omega$. In the linear response regime, the expansion can be truncated at second order, and free energy at temperature $T$ is given by~\cite{LandauLifshitz}
\begin{equation}
  F(\Omega) = F +\Omega \langle \hat R \rangle - \tfrac{1}{2} \chi \Omega^2.  \label{F_perturbation}
\end{equation}
Here, all the thermodynamic quantities on the right hand side of Eq.~(\ref{F_perturbation}) are for the equilibrium state without Raman coupling at same temperature: $F$ is free energy, $\langle \hat R \rangle $ is the ensemble average value of $\hat R$, and $\chi$ is referred to as Raman susceptibility with a formalistic expression
\begin{equation}
  \chi= 2\sum_{\ell \neq \ell'} \rho_\ell \frac{\big| \langle \Phi_{\ell'} \mid \hat R \mid \Phi_\ell\rangle \big|^2}{E_{\ell'}-E_\ell}, \label{chi}
\end{equation}
where $|\Phi_\ell\rangle$  is the $\ell$-th eigen-state of the Hamiltonian $\hat H_0 + \hat H_{\rm int}$, $E_\ell$ is the corresponding eigen-energy,  $\rho_\ell=e^{-E_\ell/k_{\rm B}T}/Z$, and $Z$ is the partition function. At $T=0$, Eqs.~(\ref{F_perturbation}) and (\ref{chi}) just reduce to the usual second order perturbation formula in quantum mechanics.

The expansion of free energy in Eq.~(\ref{F_perturbation}) can be further simplified by recognizing that the average value of $\hat R$ always vanishes in the absence of Raman coupling, i.e.,
\begin{equation}
  \langle \hat R \rangle =0.
\end{equation}
This is because the equilibrium states in the case of $\Omega=0$ persist the translational invariance symmetry, while the operator $\hat R$ does not commute with total momentum. An alternative argument is based on the fact that the free energy of the STR/PW phase should not depend on the sign of $\Omega$; hence the linear term in the expansion must vanish.

We note that the effective Hamiltonian~(\ref{Hamiltonian-1}) is written in the laboratory frame.  In contrast, the rotating frame is frequently used in previous studies~\cite{Ho,Trento1,Trento2,Trento3,Trento4,ZhengWei}, where a unitary transformation $\hat U=e^{ik_{\rm r}x\check\sigma_z}$ is performed. While the rotating frame is useful in many cases, the laboratory frame is much more convenient to our problem, because all the quantities in the perturbation theory only concern the states without Raman coupling. One can readily check that the value of the Raman susceptibility $\chi$ is actually independent of the frame~\cite{note_frame}.

A key ingredient in the perturbation theory is that the equilibrium states evolve smoothly when Raman coupling is switched on. To shed more light on this point, it is helpful to recall the condensate wave function at $T=0$. Consider the symmetric case with $g_{\uparrow\uparrow}=g_{\downarrow\downarrow}=g$ and $h=0$. In the rotating frame, the STR phase and the PW phase can be described by a variational wave function~\cite{Trento1},
\begin{equation}
  \tilde \varphi = \sqrt{n}\left[c_+ \begin{pmatrix} \cos\eta \\ -\sin\eta \end{pmatrix} e^{i k_0 x} + c_- \begin{pmatrix} \sin\eta \\ -\cos\eta \end{pmatrix} e^{-i k_0 x}\right], \nonumber
\end{equation}
with $n=N/V$ the total density of atoms. In the STR phase, $|c_+|^2=|c_-|^2=1/2$; in the PW phase, one of the coefficients $c_\pm$ is zero. For a given Raman strength $\Omega$, variation parameters $\eta$ and $k_0$ have been determined in Ref~\cite{Trento1}. As $\Omega\rightarrow 0$, one finds $k_0\rightarrow k_{\rm r}$ and $\eta\rightarrow 0$ in both the STR phase and the PW phase, therefore, the condensate wave functions in the laboratory frame $(\varphi=\hat U^{-1}\tilde \varphi)$ reduce to
\begin{gather}
  \varphi_{\rm STR} \xrightarrow{\Omega\rightarrow 0}  \sqrt{\tfrac{n}{2}} \begin{pmatrix} e^{i\theta_\uparrow} \\ e^{i\theta_\downarrow} \end{pmatrix},
  \label{wf1} \\
  \varphi_{\rm PW}  \xrightarrow{\Omega\rightarrow 0}  \sqrt{n} \begin{pmatrix} e^{i\theta_\uparrow} \\ 0 \end{pmatrix} \quad \text{or} \quad
  \sqrt{n} \begin{pmatrix} 0 \\ e^{i\theta_\downarrow} \end{pmatrix}, \label{wf2}
\end{gather}
with $\theta_\sigma$ the phase of $\sigma$-component. As is well known, the unpolarized BEC and ferromagnetic BEC described by Eqs.~(\ref{wf1}) and (\ref{wf2}) are two possible ground states in a spin-$\tfrac{1}{2}$ Bose gas without interspecies coupling. When the Raman coupling is gradually switched on, these two phases continuously evolve into the STR phase and the PW phase, respectively. At finite temperature, although equilibrium states are not characterized by the condensate wave-function alone, similar connections are still expected.

In the following, we apply the perturbation formalism to study the transition between the STR phase and the PW phase. For this purpose, we focus on the case with $g_{\uparrow\uparrow}=g_{\downarrow\downarrow}=g$ and $h=0$, where the Hamiltonian possesses a $Z_2$ symmetry~\cite{note1}. And all the interactions are assumed to be repulsive. This simplified Hamiltonian is a minimal model to understand the ground state phase diagram~\cite{Ho,Trento1}, and it is also relevant to the available experiments with rubidium atoms~\cite{NIST,ChenShuai}.  The extension to more complicated situations is straightforward.

\section{Thermodynamics in absence of Raman coupling}

In this section, we study the thermodynamics of two-component Bose gases without Raman coupling. For the convenience of later discussions,  we use $E_{\rm r}=\hbar^2 k_{\rm r}^2/(2m)$ as the energy unit in the numerical calculation. The density of atoms $n$ and recoil wave-vector $k_{\rm r}$ are set by the typical values in experiments~\cite{NIST,ChenShuai}.

\subsection{Popov approximation}

In a BEC state, condensate and non-condensed atoms can be treated separately. In our case, the field operator can be written as
\begin{equation}
  \hat \psi_\sigma = \varphi_\sigma + \delta \hat \psi_\sigma = e^{i\theta_\sigma} \Big( \sqrt{n_{0\sigma}} + \tfrac{1}{\sqrt{V}} {\sum_{\bf p}}'\hat\psi_{\bf p\sigma} e^{i{\bf p \cdot r}/\hbar} \Big)
\end{equation}
where $\varphi_\sigma=\langle \hat\psi\rangle $ is the condensate wave function of the $\sigma$-component, $n_{0\sigma}$ is the condensate density, and $\sum'_{\bf p}$ denotes the summation excluding zero momentum. Non-condensed atoms, which are usually negligible in the ground state, may play an important role at higher temperatures due to thermal fluctuations. In weakly interacting gases, the interactions between non-condensed atoms can be treated in the mean-field manner. A widely used mean-field prescription is the Popov approximation~\cite{Popov,Griffin}, which recovers the Bogoliubov theory in low temperature limit and reduces to the Hartree-Fock theory when the condensate vanishes above $T_{\rm c}$. In the Popov approximation, the grand-canonical Hamiltonian of a spin-$\frac{1}{2}$ Bose gas is given by
$$\hat K = K^{(0)} + \hat K^{(2)},$$
with $K^{(0)}=V[-\mu n_0 +g(\tfrac{1}{2} n_{0\uparrow}^2 + \tfrac{1}{2} n_{0\downarrow}^2 - \delta n_{\uparrow}^2 - \delta n_{\downarrow}^2) + g_{\uparrow\downarrow} (n_{0\uparrow} n_{0\downarrow} -\delta n_\uparrow \delta n_\downarrow -\delta s^2) ]$ and
\begin{align}
   & \hat K^{(2)} =   {\sum_{\bf p}}'\sum_{\sigma=\uparrow,\downarrow} \Big[ (\xi_{\bf p}+2g n_\sigma + g_{\uparrow\downarrow} n_{\bar \sigma}) \hat \psi_{\bf p\sigma}^\dagger \hat \psi_{\bf p\sigma} \nonumber \\
   & \qquad   + \tfrac{1}{2} \big( gn_{0\sigma} \hat \psi_{\bf p\sigma}^\dagger \hat \psi_{-\bf p\sigma}^\dagger + g_{\uparrow\downarrow} \sqrt{n_{0\uparrow}n_{0\downarrow}} \hat\psi_{\bf p\sigma}^\dagger \hat \psi_{-\bf p\bar\sigma}^\dagger  + {\rm H.c.} \big) \nonumber \\
   & \qquad  +  g_{\uparrow\downarrow} ( \sqrt{n_{0\uparrow}n_{0\downarrow}} + \delta s ) \hat\psi_{\bf p\sigma}^\dagger \hat \psi_{\bf p\bar\sigma} \Big]. \label{Popov}
\end{align}
Here, $\delta n_\sigma=\tfrac{1}{V}\sum_{\bf p}' \langle \hat\psi_{\bf p\sigma}^\dagger \hat\psi_{\bf p\sigma} \rangle$ and $n_\sigma=n_{0\sigma}+\delta n_\sigma$ are the non-condensate density and total density of the $\sigma$-component respectively, $\delta s=\tfrac{1}{V}\sum_{\bf p}' \langle \hat\psi_{\bf p\uparrow}^\dagger \hat\psi_{\bf p\downarrow} \rangle $ is the spin-flipping mean-field parameter, $n_0=n_{0\uparrow}+n_{0\downarrow}$ is total condensate density, $\mu$ is chemical potential, $\xi_{\bf p}=\epsilon_{\bf p}-\mu=p^2/(2m)-\mu$, and $\bar\sigma$ denotes the spin opposite to $\sigma$.

The condensate wave function should be determined by energy minimization. The stationary conditions $\langle \partial \hat K / \partial \varphi_\sigma \rangle =0$ can be explicitly written as
\begin{gather}
  \mu \varphi = \mathcal{L} \varphi,  \label{stationary}
\end{gather}
with $\varphi =\big( \begin{smallmatrix} \varphi_\uparrow \\ \varphi_\downarrow \end{smallmatrix} \big)$ and
\begin{align}
  \mathcal{L} = \begin{pmatrix}
    g(n_{0\uparrow}+2\delta n_\uparrow) + g_{\uparrow\downarrow} n_\downarrow & g_{\uparrow\downarrow} \delta s\, e^{i(\theta_\uparrow-\theta_\downarrow)} \\ g_{\uparrow\downarrow} \delta s \,e^{i(\theta_\downarrow-\theta_\uparrow)} & g(n_{0\downarrow}+2\delta n_\downarrow) + g_{\uparrow\downarrow} n_\uparrow
  \end{pmatrix}. \nonumber
\end{align}
At low temperature, where non-condensed atoms can be safely ignored, Eq.~(\ref{stationary}) reduces to the time-independent Gross-Pitaevskii equation. Once the condensate wave function is determined, the quadratic Hamiltonian can be readily solved via a Bogoliubov transformation.

\subsection{Equation of state}

There are two possible equilibrium states satisfying the stationary equation (\ref{stationary}). One is the spin-balanced phase, where both condensate and non-condensed atoms are unpolarized, i.e., $n_{0\uparrow}=n_{0\downarrow}$ and $\delta n_\uparrow = \delta n_\downarrow$. Through a standard diagonalization procedure, we obtain the free energy of the spin-balanced phase as
\begin{equation}
  F_{\rm B} = E_{\rm B0} + k_{\rm B} T {\sum_{\bf p}}' \sum_{\alpha=\pm} \ln \big(1-e^{-\hbar \omega_{{\bf p}\alpha}/ k_{\rm B}T}\big), \label{F_balanced}
\end{equation}
where $ E_{\rm B0} = V [\tfrac{1}{4}(g_+n^2+g\delta n^2)+g_{\uparrow\downarrow}\delta s (\delta n-\delta s)] + \tfrac{1}{2}\sum_{\bf p}'[ \hbar \omega_{p+} + \hbar\omega_{p-}-2\epsilon_p-gn_0+2g_{\uparrow\downarrow}\delta s + (g^2+g_{\uparrow\downarrow}^2)n_0^2/(4\epsilon_p)]$ with $\delta n =\delta n_\uparrow + \delta n_\downarrow$, and $\hbar \omega_{\bf p\pm}$ are excitation spectra of quasi-particles,
\begin{gather}
  \hbar \omega_{\bf p+} = \sqrt{\epsilon_p(\epsilon_p + g_+n_0)}, \\
  \hbar \omega_{\bf p-} = \sqrt{(\epsilon_p-2g_{\uparrow\downarrow}\delta s)(\epsilon_p-2g_{\uparrow\downarrow}\delta s + g_-n_0)},
\end{gather}
with $g_\pm \equiv g\pm g_{\uparrow\downarrow}$. The mean-field parameters $\delta n$ and $\delta s$ should be self-consistently determined from
\begin{gather}
  \delta n =  \frac{2}{V} {\sum_{\bf p}}' \sum_{\alpha=\pm} \left(u_{\bf p\alpha}^2+v_{\bf p\alpha}^2 \right)f_{\bf p\alpha}, \label{selfconsist1}\\
  \delta s = \frac{1}{V}  {\sum_{\bf p}}' \sum_{\alpha=\pm} \alpha \left(u_{\bf p\alpha}^2+v_{\bf p\alpha}^2 \right)f_{\bf p\alpha}, \label{selfconsist2}
\end{gather}
where $f_{\bf p\alpha}=1/(e^{\hbar \omega_{\bf p\alpha} / k_{\rm B}T}-1)$, $u_{\bf p,\alpha}$ and $v_{\bf p,\alpha}$ are coefficients of the Bogoliubov transformation with $u_{\bf p,\alpha}^2=v_{\bf p,\alpha}^2+\tfrac{1}{2}$, $v_{{\bf p}+}^2 = \tfrac{1}{8}[ \sqrt{(\epsilon_p+g_+n_0) / (\hbar\omega_{{\bf p}+})} - \sqrt{\epsilon_p / (\hbar\omega_{{\bf p}+})} ]^2 $, and $v_{{\bf p}-}^2 = \tfrac{1}{8} [ \sqrt{(\epsilon_p-2g_{\uparrow\downarrow}\delta s+g_-n_0)/ (\hbar \omega_{{\bf p}-})} - \sqrt{(\epsilon_p-2g_{\uparrow\downarrow}\delta s) / (\hbar \omega_{{\bf p}-})} ]^2 $.
In Eqs.~(\ref{selfconsist1}) and (\ref{selfconsist2}), we have neglected the terms associated with quantum depletions at zero temperature. This treatment is well justified in dilute gases. Obviously, only a negative $\delta s$ is allowed in the self-consistency equations.

Intuitively, the spin-balanced phase is favored in the case $g_->0$, where the intra-species repulsion is stronger. For $g_-<0$, the excitation branch $\hbar\omega_{{\bf p},-}$ suffers a dynamic instability in the low temperature limit, and the spin-balanced phase is not available until temperature beyond a threshold value. The threshold temperature $T_{\rm B}$ is determined by the stability condition
\begin{equation}
  g_-n_0 - 2g_{\uparrow\downarrow}\delta s\geqslant 0,
\end{equation}
when the equality is satisfied. Above $T_{\rm B}$, the spin-balanced phase becomes a (meta)stable state corresponding to a (local) minimum in the free energy landscape.

Another possible equilibrium state is the spin-polarized phase, which spontaneously breaks $Z_2$ symmetry. In this phase, the condensate is fully polarized, and spin-up and spin-down are decoupled ($\delta s=0$). The free energy of the spin-polarized phase (assume $M_0=+1$) is given by
\begin{equation}
  F_{\rm P}= E_{\rm P0} +  k_{\rm B} T {\sum_{\bf p}}' \sum_{\sigma=\uparrow,\downarrow} \ln \big(1-e^{-\hbar \omega_{{\bf p}\sigma} / k_{\rm B}T} \big), \label{F_polarized}
\end{equation}
where $ E_{\rm P0} = V[\tfrac{1}{2}gn^2 + \tfrac{1}{4} g_{\uparrow\downarrow}\delta n^2 + (g-\tfrac{1}{2}g_{\uparrow\downarrow}) \delta n^2 \delta M (1-\tfrac{1}{2}\delta M) + \tfrac{8}{15} gn_0^2 (mgn_0)^{3/2}/(\pi^2 \hbar^3 n_0)]$ with $\delta M= (\delta n_\uparrow-\delta n_\downarrow)/\delta n$, and $\hbar\omega_{\bf p\sigma}$ are the excitation spectra of quasi-particles
\begin{gather}
  \hbar \omega_{\bf p\uparrow} = \sqrt{\epsilon_p(\epsilon_p+2gn_0)}, \\
  \hbar \omega_{\bf p\downarrow} = \epsilon_p - g_-n_0 - (2g-g_{\uparrow\downarrow}) \delta n\delta M.
\end{gather}
The mean-field parameters $\delta n_\sigma$ should be determined by the self-consistency equations
\begin{align}
  \delta n &=  \frac{1}{V} {\sum_{\bf p}}' \left[(u_{\bf p}^2+v_{\bf p}^2)f_{\bf p\uparrow}+ f_{\bf p\downarrow}\right], \\
  \delta n\delta M &= \frac{1}{V} {\sum_{\bf p}}' \left[(u_{\bf p}^2+v_{\bf p}^2)f_{\bf p\uparrow}- f_{\bf p\downarrow}\right],
\end{align}
where $f_{\bf p\sigma}=1/(e^{\hbar\omega_{p\sigma}/k_{\rm B}T}-1)$, and $u_{\bf p}^2=v_{\bf p}^2+1=\tfrac{1}{2}[(\epsilon_p+gn_0)/(\hbar\omega_{p\uparrow})+1]$.

For $g_->0$, the excitation branch $\hbar \omega_{\bf p\downarrow}$ suffers an energetic instability in the low temperature limit, and the spin-polarized phase is only available when the stability condition
\begin{equation}
  g_-n_0 + (2g-g_{\uparrow\downarrow}) \delta n \delta M \leqslant 0
\end{equation}
is satisfied. The equality of above condition determines a threshold temperature $T_{\rm P}$. Above $T_{\rm P}$, $\hbar \omega_{\bf p\downarrow}$ is gapped, and the spin-polarized phase becomes a (meta)stable equilibrium state. For $g_-\ll g$, $T_{\rm P}$ is indeed very low.

\begin{figure}[t!]
\includegraphics[width=6.9cm]{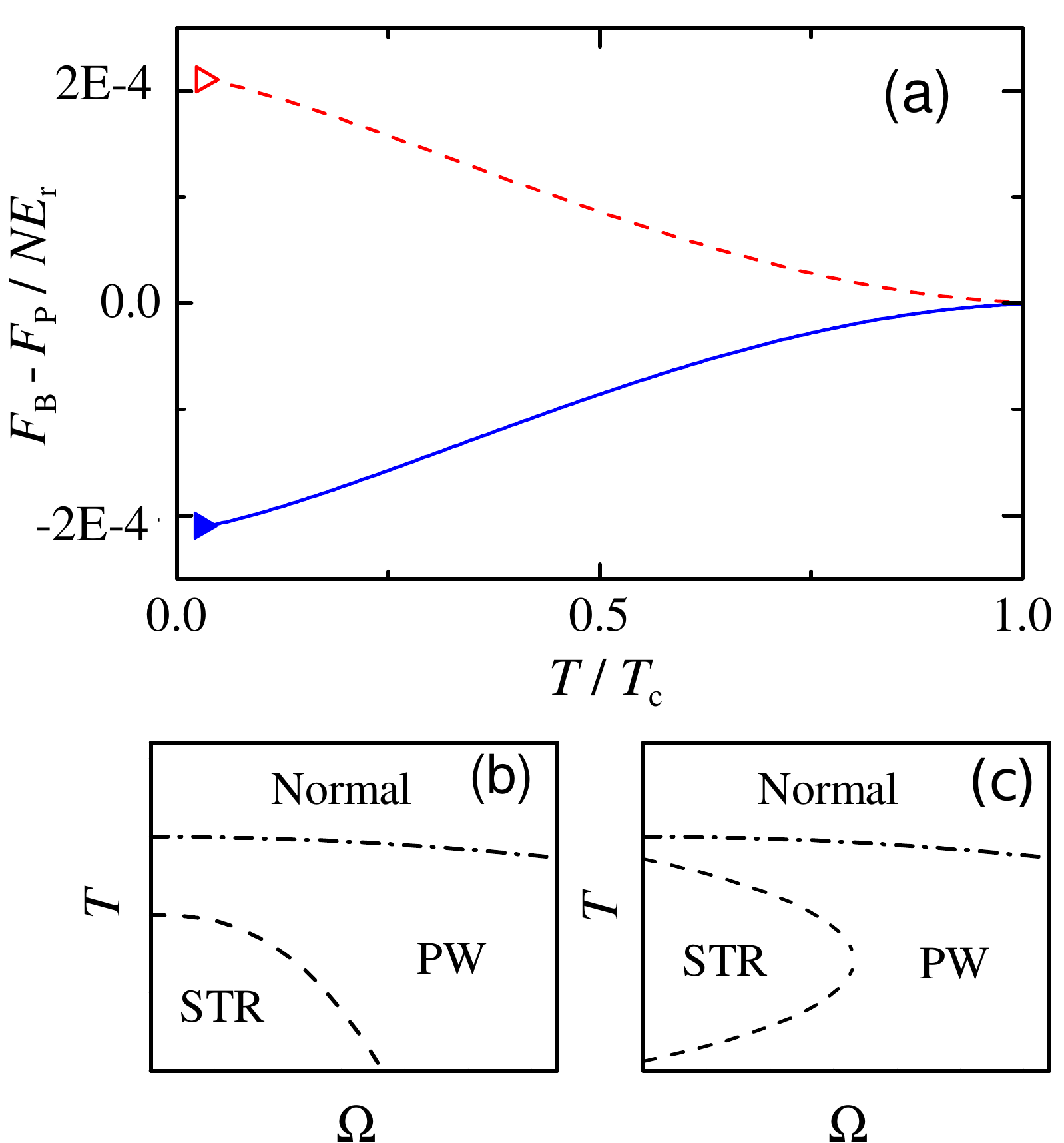}
\caption{(color online). (a) Comparison of the free energy of the spin-balanced phase and the spin-polarized phase at finite temperature for $g_-/g=0.002$ (solid line) and $g_-/g=-0.002$ (dashed line). The threshold temperatures for the spin-balanced phase and the spin-polarized phase are indicated by $\triangleright$ and $\blacktriangleright$, respectively (see text). Parameters: $n=0.5 k_{\rm r}^3$, $g=0.828 E_{\rm r}/k_{\rm r}^3$. $T_{\rm c}=2\pi \hbar^2 [n/\zeta(\tfrac{3}{2})]^{2/3}/(mk_{\rm B})$ is the transition temperature in the noninteracting case with $\zeta(\cdot)$ the Riemann zeta function. Lower panels show two ruled-out scenarios of the phase diagram with Raman coupling: (b) a ruled out scenario  for $g_->0$, and (c) a ruled out scenario for $g_-<0$.} \label{fig-1}
\end{figure}

In Fig.~1, we numerically compare the free energy of the spin-balanced phase and the spin-polarized phase at finite temperature. For $g_->0$, the spin-balanced phase is the only possible equilibrium state when temperature is below the threshold value $T_{\rm p}$. Although the spin-polarized phase becomes available at higher temperature, a free-energy crossing is never observed; hence the system prefers the spin-balanced phase up to the condensation temperature $T_{\rm c}$. On the other hand, for $g_-<0$, $F_{\rm P}<F_{\rm B}$ holds in the entire temperature region $T_{\rm B}<T<T_{\rm c}$, which means the spin-polarized phase is  energetically  favored in this case.

The above results about equation of state provide strong constraints for the phase diagram in the presence of Raman coupling. Since there is a finite free energy difference between the spin-balanced phase and the spin-polarized phase, an infinitesimal Raman coupling could not induce any phase transition. Therefore, for $g_->0$, the STR-PW transition in the weak Raman coupling limit is impossible unless temperature approaches $T_{\rm c}$, i.e., the scenario of the phase diagram as shown in Fig.~1(b) can be completely ruled out~\cite{note_phasediagram}.
Similarly, for $g_-<0$, the STR phase at finite temperature can be also excluded in  the weak Raman coupling regime [see Fig.~1(c)].

\subsection{Raman susceptibility}

According to Kubo's formula, the fluctuation $\delta \langle\hat R\rangle$ generated by the Raman perturbation is described by the dynamic response function~\cite{KuboFormula},
\begin{equation}
  \chi_{\rm R}(\omega) = \frac{i}{\hbar} \int_0^\infty dt\, \langle \big[ \hat R(t), \hat R \big] \rangle e^{i\omega t},
\end{equation}
where  the time-dependent operator $\hat R(t)$ is defined in the Heisenberg picture as usual. Using the Lehmann representation, one can immediately recognize that the Raman susceptibility in Eq.~(\ref{chi}) is just the static response, $\chi=\chi_{\rm R}(\omega=0)$.

As a leading order approximation, we ignore the dynamics of non-condensate mean-fields  and write the time-dependent field operator as
\begin{align}
  \hat \psi_\sigma(t) &= e^{i(\theta_\sigma-\mu t/\hbar)} \Big[ \sqrt{n_{0\sigma}} \nonumber \\
  & \qquad + \tfrac{1}{\sqrt{V}} {\sum_{\bf p}}'e^{i\hat Kt / \hbar} \hat \psi_{\bf p\sigma} e^{-i\hat Kt / \hbar} e^{i \bf p\cdot r/\hbar} \Big],
\end{align}
where $\hat K$ is the static Popov Hamiltonian given by Eq.~(\ref{Popov}). Since $\hat K$ is diagonal in quasi-particles representation, $\chi_{\rm R}$ can be easily derived.
After straightforward algebra, we obtain Raman susceptibility in the spin-balanced phase and the spin-polarized phase as
\begin{widetext}
\begin{align}
  \chi_{\rm B} &= N_0 \sum_{\alpha=\pm} \frac{(u_{2{\bf k}_{\rm r}\alpha} - \alpha\, v_{2{\bf k}_{\rm r}\alpha})^2}{2 \hbar \omega_{2{\bf k}_{\rm r}\alpha}} + {\sum_{\bf p}}'' \sum_{\alpha\alpha'=\pm}  \Big[ \frac{(u_{{\bf p}+2{\bf k}_{\rm r}\alpha} v_{\bf p\alpha'} + \alpha\alpha' u_{\bf p\alpha'} v_{{\bf p}+2{\bf k}_{\rm r}\alpha})^2(f_{{\bf p}\alpha'}+ f_{{\bf p}+2{\bf k}_{\rm r}\alpha})}{2\hbar(\omega_{{\bf p}\alpha'}+\omega_{{\bf p}+2{\bf k}_{\rm r}\alpha})} \nonumber  \\
    & \qquad \qquad \qquad - \frac{(u_{{\bf p}+2{\bf k}_{\rm r}\alpha} u_{\bf p\alpha'} + \alpha\alpha' v_{\bf p\alpha'} v_{{\bf p}+2{\bf k}_{\rm r}\alpha})^2(f_{{\bf p}\alpha'}- f_{{\bf p}+2{\bf k}_{\rm r}\alpha})}{2\hbar(\omega_{{\bf p}\alpha'}-\omega_{{\bf p}+2{\bf k}_{\rm r}\alpha})} \Big], \label{chi_balanced}
   \\
   \chi_{\rm P} &= \frac{N_0}{2\hbar\omega_{2{\bf k}_{\rm r}\downarrow}}  + \frac{1}{2}{\sum_{\bf p}}' \Big[ v_{\bf p}^2 \frac{f_{\bf p\uparrow} + f_{{\bf p}+2{\bf k}_{\rm r}\downarrow}}{\hbar(\omega_{\bf p\uparrow}+\omega_{{\bf p}+2{\bf k}_{\rm r}\downarrow})} -u_{\bf p}^2 \frac{f_{\bf p\uparrow} - f_{{\bf p}+2{\bf k}_{\rm r}\downarrow}}{\hbar(\omega_{\bf p\uparrow}-\omega_{{\bf p}+2{\bf k}_{\rm r}\downarrow})}  \Big],  \label{chi_polarized}
\end{align}
\end{widetext}
where ${\bf k}_{\rm r}=(k_{\rm r},0,0)$, and $\sum_{\bf p}''$ denotes a summation with the constraint ${\bf p}\neq 0,-2\hbar{\bf k}_{\rm r}$. At low temperature, the contribution from thermal atoms is negligible, Raman susceptibility can be written in analytical forms,
\begin{align}
  \chi_{\rm B} &= \frac{N}{16E_{\rm r}}+ \frac{N}{16E_{\rm r}+8g_+n}, \\
  \chi_{\rm P} &= \frac{N}{8E_{\rm r}-2g_-n},
\end{align}
where we have replaced $N_0$ by the total number of atoms.
In the weak interacting limit, both Eqs.~(\ref{chi_balanced}) and (\ref{chi_polarized}) approach the noninteracting result (see Appendix A).

In Fig.~2, we plot Raman susceptibility $\chi_{\rm B}$ and $\chi_{\rm P}$ as a function of temperature. In spite of a non-monotonic temperature dependent behavior, $\chi_{\rm B}$ is always smaller than $\chi_{\rm P}$ for $T<T_{\rm c}$. This fact implies free energy of the spin-polarized phase will decrease faster when  Raman coupling is switched on [see Eq.~(\ref{F_perturbation})]. Therefore, for $g_->0$, a transition between the STR phase and the PW phase is expected at a critical Raman coupling strength.

\begin{figure}[b]
\includegraphics[height=5cm]{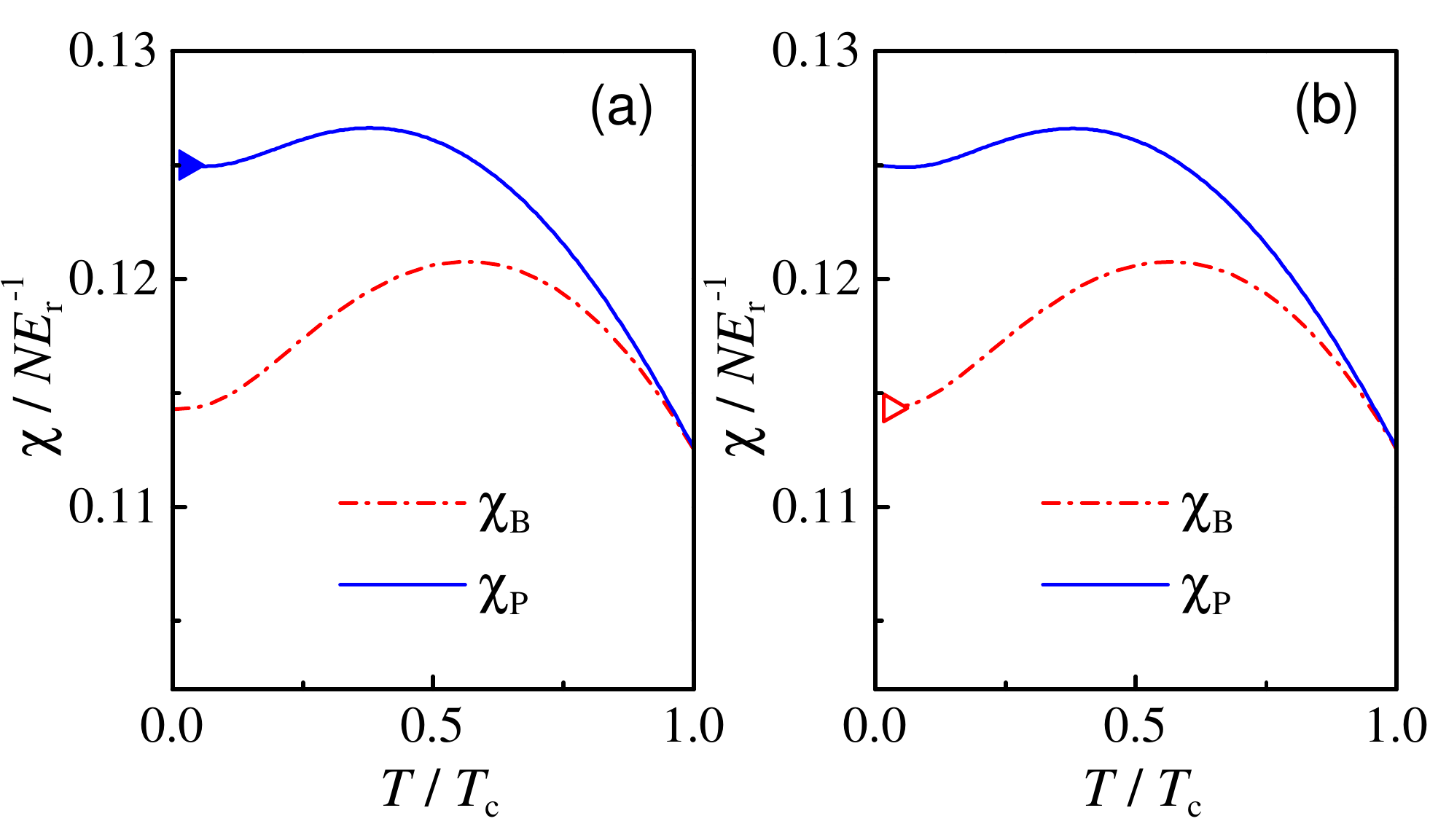}
\caption{(color online). Raman susceptibility $\chi$ of the spin-balanced phase and the spin-polarized phase at finite temperature for (a) $g_-/g=0.002$ and (b) $g_-/g=-0.002$. The threshold temperatures for the spin-balanced phase and the spin-polarized phase are indicated by $\triangleright$ and $\blacktriangleright$, respectively. Other parameters are the same as in Fig.~1.}\label{fig-2}
\end{figure}

\section{Transition between stripe phase and plane-wave phase}

\subsection{Critical Raman strength}

With the equation of state and Raman susceptibility obtained previously, we can determine the phase diagram in the presence of Raman coupling via the perturbation approach. According to Eq.~(\ref{F_perturbation}), the free energies of the STR phase and the PW phase are given by
\begin{eqnarray}
  F_{\rm STR}(\Omega) &=& F_{\rm B} - \tfrac{1}{2}\chi_{\rm B} \,\Omega^2 , \\
  F_{\rm PW}(\Omega) &=& F_{\rm P} - \tfrac{1}{2}\chi_{\rm P} \,\Omega^2.
\end{eqnarray}
For the case $g_->0$, $F_{\rm B}<F_{\rm P}$; hence the system is in the STR phase when Raman coupling is weak enough. As $\Omega$ increases, a first order transition takes place when the condition $F_{\rm STR}(\Omega_{\rm c})= F_{\rm PW}(\Omega_{\rm c})$ is satisfied, and the critical Raman strength can be explicitly written as
\begin{equation}
  \Omega_{\rm c}^2 = 2\frac{F_{\rm B}-F_{\rm P}}{\chi_{\rm B}-\chi_{\rm P}} \label{phase_boundary}.
\end{equation}
When $\Omega>\Omega_{\rm c}$, the PW phase is energetically favored.
For small (positive) $g_-$, the critical Raman strength $\Omega_{\rm c}$ is also small because the free energies of the spin-balanced phase and the spin-polarized phase are very close. This is indeed  the case in current experiments with Rb atoms~\cite{NIST,ChenShuai}.  We note that since the spin-polarized phase is not available at very low temperature,  the phase transition can be only addressed above the threshold temperature $T_{\rm P}$, which is a limitation of the perturbation approach.

In Fig.~3, the phase boundary obtained from Eq.~(\ref{phase_boundary}) is plotted for various densities. As temperature increases, the phase boundary bends toward the stripe phase side in most temperature regions, which implies the PW phase is more robust than the STR phase in the presence of thermal fluctuations.  At higher temperature close to $T_{\rm c}$~\cite{note_Tc}, the critical Raman strength shows a suspicious non-monotonic behavior. Since both $F_{\rm B}-F_{\rm P}$ and $\chi_{\rm B}-\chi_{\rm P}$ vanish at $T_{\rm c}$, the value of $\Omega_{\rm c}$ is sensitive to the temperature dependence details of all quantities. In fact, if we use the Hartree-Fock approximation to compute $F$ and $\chi$ (see Appendix B), the phase boundary shows a quite different behavior in the vicinity of $T_{\rm c}$. As is well known, mean-field theories usually produce artificial results near $T_{\rm c}$~\cite{near_Tc}; thus the STR-PW transition in this narrow region is not conclusive.

At $\Omega_{\rm c}$, we also numerically check the inhomogeneous state with a spacial separation between the STR phase and the PW phase. The density jump across the interface is found to be extremely small, hence the phase separation is almost invisible in a uniform system. Previously, the variational study at zero temperature came to a similar conclusion~\cite{Trento2}.

\begin{figure}
\includegraphics[height=5cm]{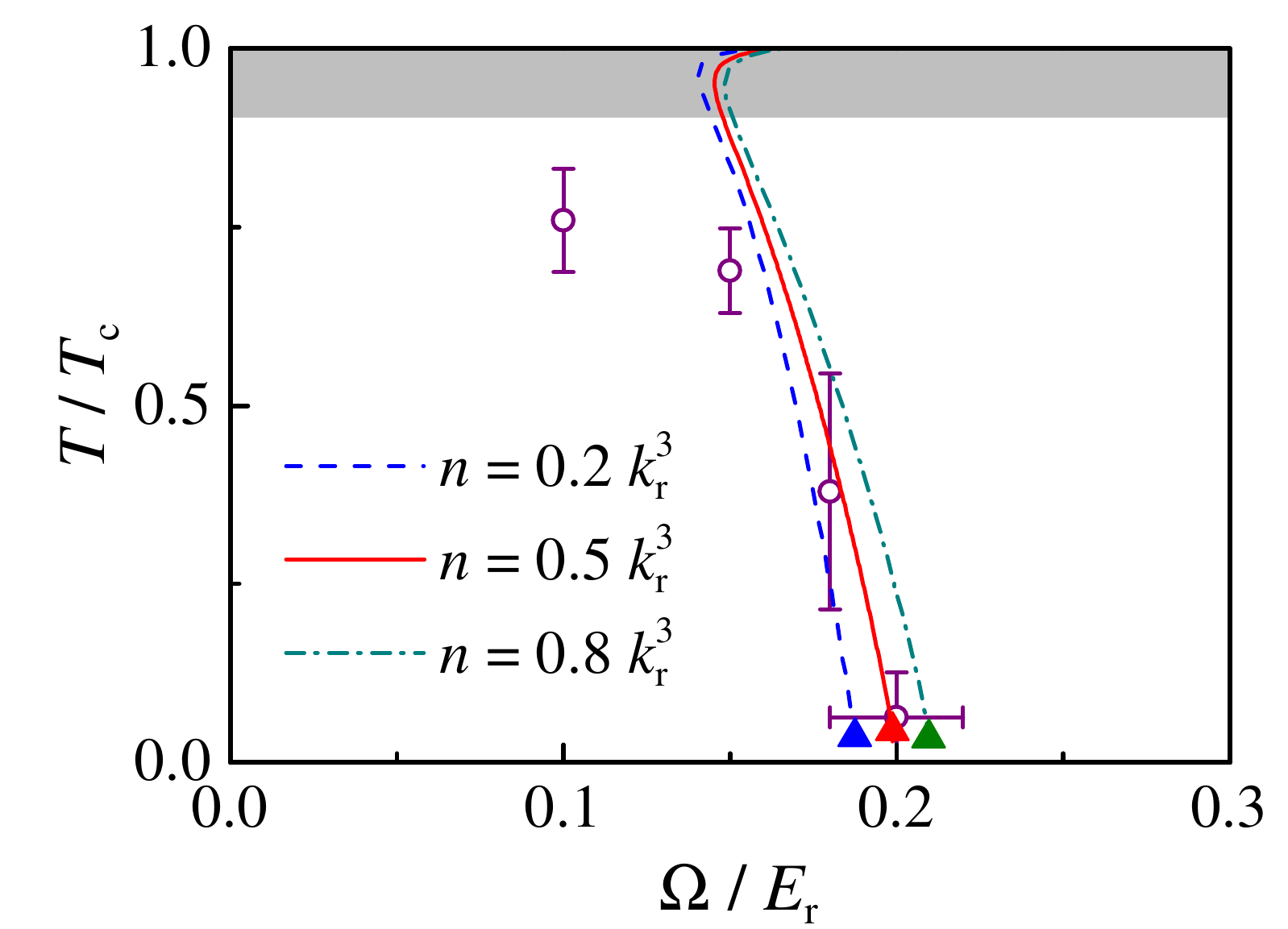}
\caption{(color online). The transition lines between the STR phase and the PW phase for various densities. $g_-/g=0.002$; other parameters are the same as in Figs.~1 and 2. The threshold temperature for the spin-polarized phase is indicated by $\blacktriangle$. When temperature is close to $T_{\rm c}$ (for instance, in the shadow region above 0.9 $T_{\rm c}$), mean-field theory is usually inaccurate. For a qualitative comparison, the experimental data ($\circ$ with error bar) measured in a harmonic trap from Ref.~\cite{ChenShuai} are also shown. At low temperature, the typical atomic density at the center of the trap is around $0.5 k_{\rm r}^3$. }\label{fig-3}
\end{figure}

\subsection{Comparison with experimental measurement}

Although our calculation is performed in the uniform case, the theoretical results qualitatively agree with the recent experimental measurement in Rb gases~\cite{ChenShuai}, where the critical Raman strength $\Omega_{\rm c}$ is found moving to a smaller value as temperature increases. Particularly, for  typical atomic densities near the center of the harmonic trap~\cite{note_experiment}, the phase boundary determined from the perturbation theory is very close to the experimental results at low temperature (see Fig.~3). At higher temperature $(T\gtrsim 0.8T_{\rm c})$, there is a quantitative discrepancy between the theoretical prediction and experimental data. This discrepancy may be due to either the failure of the mean-field description or the inhomogeneous density distribution in the trap.

In the present work, only the uniform situation is considered. For the trapped case, the condensate and the thermal atoms construct an inhomogeneous shell structure. In order to determine the density profile in a harmonic trap, both the equation of state and the knowledge of the condensate fraction are needed. However, the condensate fraction at a given Raman strength $\Omega$ can not be directly obtained via the perturbation approach. At this stage, the quantitative determination of the phase diagram in a harmonic trap is still an open question, and we leave this issue to future study.

\section{Discussion and Conclusion}

The perturbation approach developed in this work is expected to be reliable when Raman coupling is weak enough. To provide an estimation of the applicable regime, we examine the equation of state in a noninteracting Bose gas with Raman coupling. As shown in Appendix A, the expansion of free energy in Eq.~(\ref{F_perturbation}) is very accurate for $\Omega\lesssim 0.2E_{\rm r}$, and the contribution from higher orders can be safely ignored in this region. A similar situation can be expected in a weakly interacting system.

The equation of state obtained via the perturbation approach can be tested by future experimental measurements and quantum Monte Carlo simulations. We note that the Raman susceptibility $\chi$ is not only a useful quantity from theoretical viewpoint but also measurable via two-photon Bragg spectroscopy. By varying the detuning of the Bragg lasers, the dynamic structure factor in the density and spin channels can be measured separately~\cite{Vale}. In the spin channel, the $f$-sum rule for the dynamic structure factor is modified by Raman coupling~\cite{note_fsum}. From the commutation relation $ \hbar^2 \int d\omega \, \omega S_{\rm M}({\bf q},\omega)=\tfrac{1}{2}\langle \big[\hat\sigma_{\bf q}^{z\dagger},[\hat H,\hat\sigma_{\bf q}^z] \big]\rangle $, we derive
\begin{equation}
  \hbar^2 \int d\omega \, \omega S_{\rm M}({\bf q},\omega) = N  \frac{\hbar^2 q^2}{2m} - 2\Omega \mathcal{R} , \label{f-sum}
\end{equation}
where $S_{\rm M}({\bf q},\omega)=\sum_{\ell,\ell'}e^{-E_\ell/k_{\rm B}T} |\langle \Phi_{\ell'}|\hat \sigma_{\bf q}^{z} | \Phi_\ell\rangle |^2 \delta(\hbar\omega-E_{\ell'}+E_\ell)$ is the spin dynamic structure factor, $\hat\sigma_{\bf q}^{z}=\int d{\bf r}\, \hat\psi^\dagger \check\sigma_z \hat \psi e^{i\bf q\cdot r}$ is the spin fluctuation operator, and $\mathcal{R}$ is the measured value of $\hat R$ in the presence of Raman coupling. The $f$-sum rule in Eq.~(\ref{f-sum}) is model-independent  and holds for both bosons and fermions.  This exact relation provides a practical way to deduce the quantity $\mathcal{R}$ through the measurement of the dynamic structure factor. Once $\mathcal{R}$ is achieved, Raman susceptibility can be readily obtained from a linear fitting: for weak Raman coupling, $\mathcal{R}$ is proportional to $\Omega$ with the simple relation $\mathcal{R}=-\chi\Omega$.

In summary, a perturbation theory of Raman coupled Bose gases is developed. The transition between the STR phase and the PW phase is investigated in the uniform case, and the phase boundary is determined at finite temperature. Our theoretical results qualitatively agree with the recent measurements in Rb gases, and the equation of state obtained here may be useful to future experiments.

\begin{acknowledgments}
  The author would like to thank helpful discussions with S. Stringari, T. Ozawa and S. Zhang. S. Chen and S.-C. Ji kindly provided the experimental data of Ref.~\cite{ChenShuai}. This work has been supported by ERC through the QGBE grant and by Provincia Autonoma di Trento.
\end{acknowledgments}

\appendix

\section{Noninteracting Bose Gas with Raman coupling}

In a non-interacting Bose gas with Raman coupling, the energy spectrum has two degenerate minima for $\Omega<4E_{\rm r}$, and there are many possibilities for the condensate to occupy these two minima. Nevertheless, the thermodynamic properties do not depend on the configuration of the condensate, and the free energy is given by
\begin{equation}
  F(\Omega)= \mu N + \sum_{\bf p}\sum_{\alpha=\pm} [\ln (1- e^{-\xi_{{\bf p},\alpha}/k_{\rm B}T}) ], \label{F_noninteracting}
\end{equation}
with $\xi_{\bf p,\pm}=\epsilon_p+E_{\rm r}\pm\sqrt{(\hbar p_xk_r/m)^2+\Omega^2/4}-\mu $. Below $T_{\rm c}$, $\mu$ equals the lowest energy of single particle dispersion. For $\Omega<4E_{\rm r}$, $\mu=-\Omega^2/(16E_{\rm r})$.

\begin{figure}
\includegraphics[height=5cm]{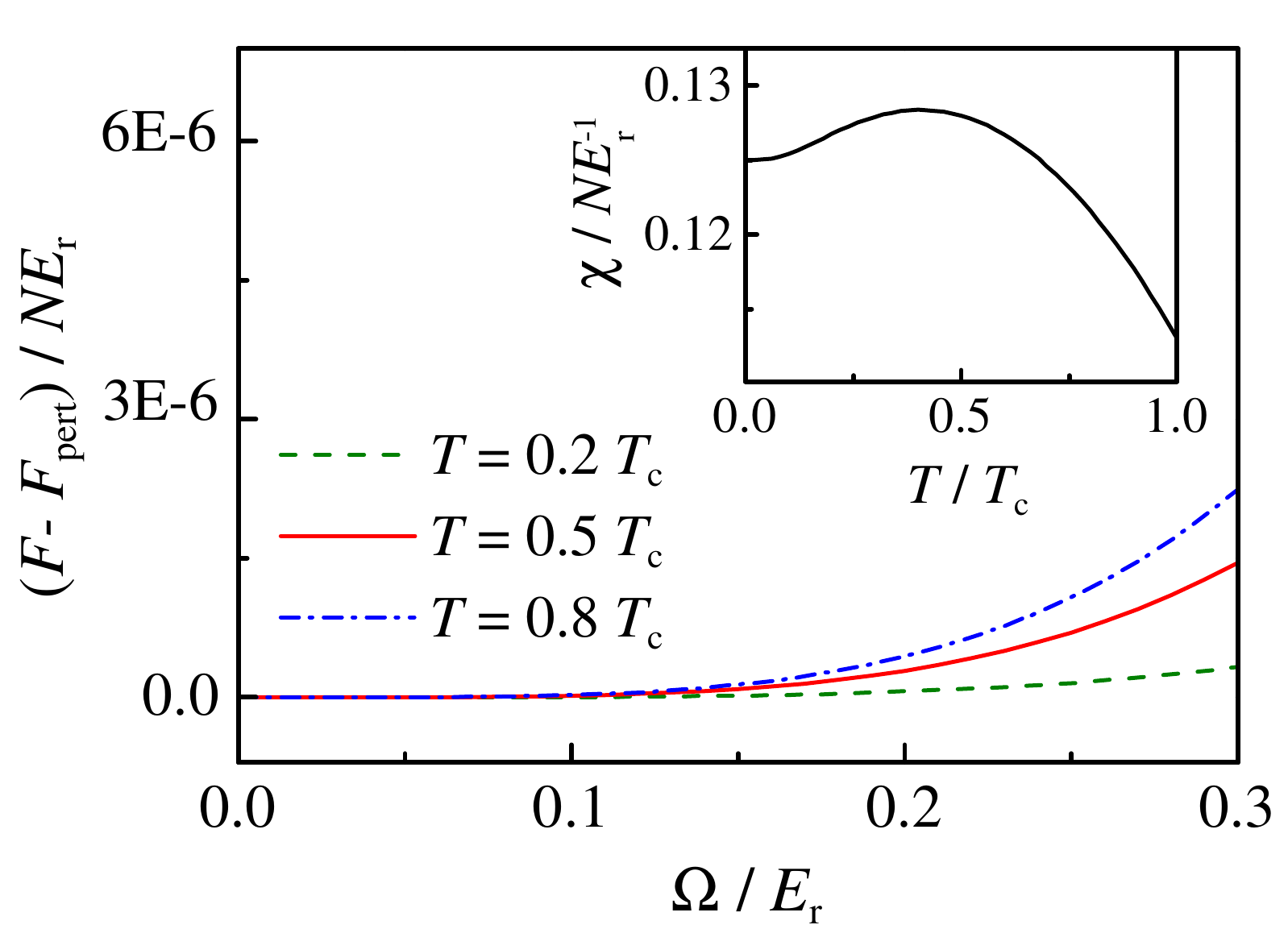}
\caption{(color online). Comparison of the actual free energy with the perturbation value $F_{\rm perp}$ in a noninteracting Bose gas at various temperatures. Inset: noninteracting Raman susceptibility as a function of temperature ($n=0.5k_{\rm r}^3$). }\label{fig-4}
\end{figure}

In Fig.~4, we numerically compare the actual free energy with the perturbation formula $F_{\rm pert}\equiv F(\Omega=0)-\tfrac{1}{2}\chi \Omega^2$, where $\chi$ is the noninteracting Raman susceptibility
\begin{equation}
  \chi=\frac{N_0}{8E_{\rm r}} +　{\sum_{\bf p}}'' \frac{m}{2\hbar k_{\rm r}(\hbar k_{\rm r}+p_x)} \frac{1}{e^{\epsilon_p/k_{\rm B}T}-1}. \label{chi_noninteracting}
\end{equation}
As one can see, for weak Raman coupling, the deviation of $F_{\rm pert}$ from the actual value is extremely small, which justifies the expansion in Eq.~(\ref{F_perturbation}) as being a very good approximation. In the inset of Fig.~4, the noninteracting Raman susceptibility is plotted as a function of temperature up to $T_{\rm c}$. \\

\section{Hartree-Fock approximation}

The grand-canonical Hamiltonian in the Hartree-Fock (HF) approximation can be readily obtained from Eq.~(\ref{Popov}) by omitting the anomalous quadratic terms. In the HF theory, the formulas for free energy and Raman susceptibility remain the same as in Popov theory except the following replacements: in the spin-balanced phase, $v_{\bf p,\alpha}\rightarrow 0$, $\hbar \omega_{\bf p+}\rightarrow \epsilon_p+\tfrac{1}{2}g_+n_0$ and $ \hbar \omega_{\bf p-}=\epsilon_p+2g_{\uparrow\downarrow}\delta s+\tfrac{1}{2}g_-n_0$; in the spin-polarized phase, $v_{\bf p}\rightarrow 0$ and $\hbar\omega_{\bf p\uparrow}\rightarrow \epsilon_p+gn_0$.

\begin{figure} [b]
\includegraphics[height=5cm]{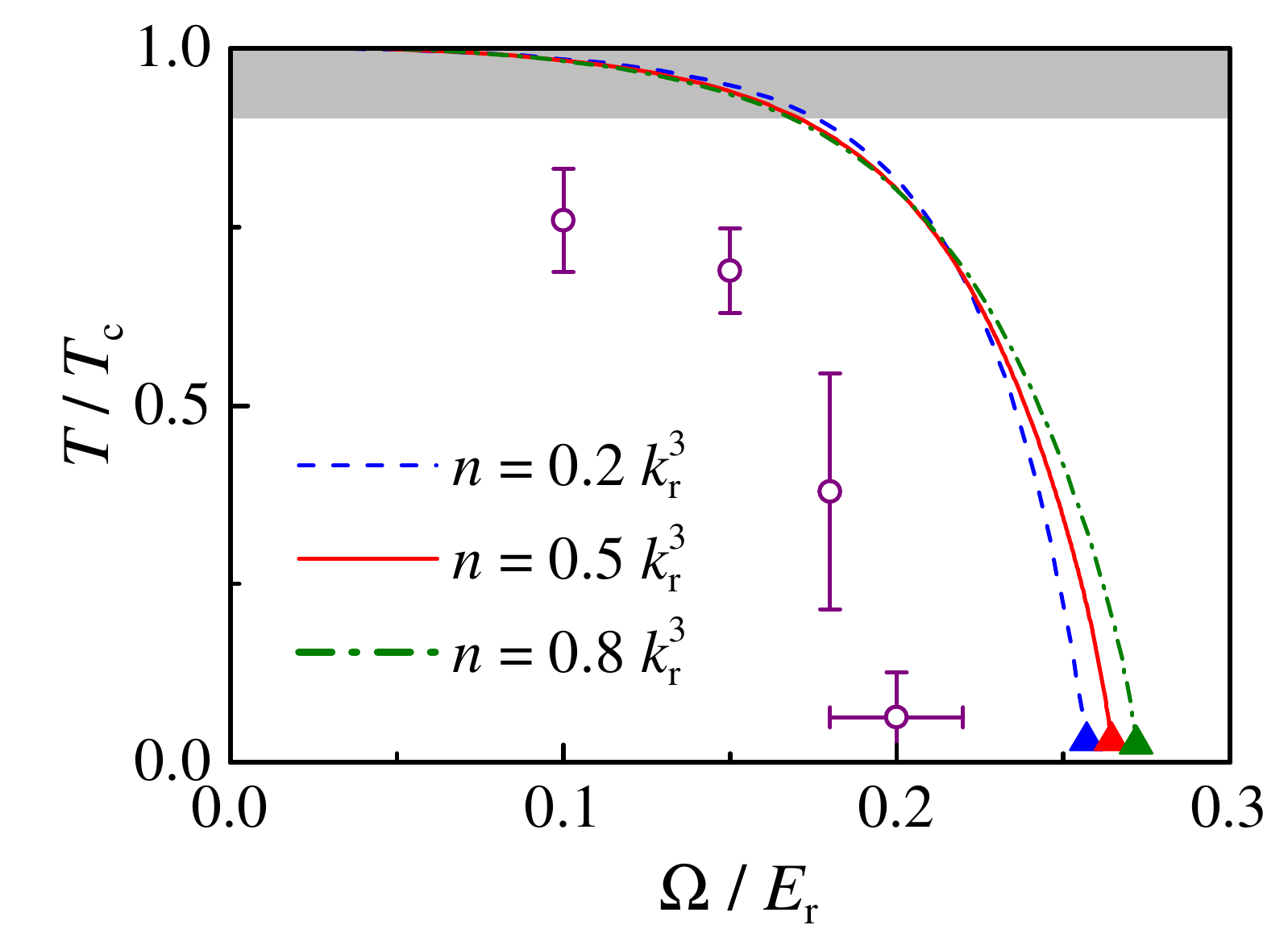}
\caption{(color online). The transition lines between the STR phase and the PW phase with $F$ and $\chi$ calculated in the Hartree-Fock approximation. Parameters and notations are the same as in Fig.~3.}\label{fig-5}
\end{figure}

In Fig.~5, the boundary between the STR phase and the PW phase is plotted with $F_{\rm B}-F_{\rm P}$ and $\chi_{\rm B}-\chi_{\rm P}$ calculated in HF approximation. At $T_{\rm c}$, the critical Raman strength $\Omega_{\rm c}$ approaches zero, which is in contrast to the results of Popov theory (see Fig.~3). As mentioned before, both Popov theory and HF theory are not reliable in the vicinity of $T_{\rm c}$. At very low temperature, where thermal fluctuations are dominated by phonons, the HF approximation is also not good due to the gapped excitation spectrum. Nevertheless, in most temperature regions the phase boundary shows a similar trend in both theories.


\begin{thebibliography}{99}
\bibitem{note_frame} In the rotating frame, $\hat R=\tfrac{1}{2}\int \hat \psi^\dagger \check \sigma_x \hat \psi$, and $\chi$ is proportional to the transverse spin susceptibility.
\bibitem{note1} In this case, the Hamiltionian is invariant under the joint operation ${\bf r}\rightarrow -{\bf r}$ and $\check\sigma_z\rightarrow -\check\sigma_{z}$. In the PW phase, this $Z_2$ symmetry is spontaneously broken.
\bibitem{note_phasediagram} However, for the case with $h\neq 0$ or $g_{\uparrow\uparrow}\neq g_{\downarrow\downarrow}$, the STR-PW transition could take place at a temperature below $T_{\rm c}$ in $\Omega\rightarrow 0$ limit.
\bibitem{note_Tc} The actual condensation temperature at finite $\Omega$ can not be obtained via the perturbation approach. Here, we set $T_{\rm c}$ as the value in the absence of Raman coupling. For small $\Omega$, the shift of $T_{\rm c}$ is indeed very small~\cite{ZhengWei}.
\bibitem{note_experiment} In the experimental work of Ref.~\cite{ChenShuai}, there is an atoms loss during the cooling procedure, which gives rise to an uncertainty of the atomic density for different temperature. In the low temperarture limit, the typical density at the center of the trap is about $n\simeq 0.5 k_{\rm r}^3$.
\end{thebibliography}
\end{document}